\documentstyle[aps,preprint]{revtex}
\draft

\begin{document}

\title{
Confined chaotic behavior in collective motion for populations of 
globally coupled chaotic elements
}
\author{Naoko NAKAGAWA}
\address{
Laboratory for Information Synthesis, 
RIKEN Brain Science Institute,
Wako, Saitama 351-01, Japan}

\author{Teruhisa S. KOMATSU}
\address{
Department of Pure and Applied Sciences, 
College of Arts and Sciences,
University of Tokyo, 
Tokyo 153, Japan}

\maketitle

\begin{abstract}
The Lyapunov exponent for collective motion is defined 
in order to characterize chaotic properties of collective motion 
for large populations of chaotic elements.
Numerical computations for this quantity suggest that 
such collective motion is always chaotic, whenever it appears.
Chaotic behavior of collective motion is found to be confined 
within a small scale,
whose size is estimated using the value of the Lyapunov exponent.
Finally, we conjecture why the collective motion appears 
low dimensional despite the actual high dimensionality of the dynamics.

\end{abstract}



\section{Introduction}

Systems consisting of large populations of interacting dynamical elements 
are widely distributed in nature,
from communities of ants to biological cell assemblies and neural networks.
Many such populations are similar in the respect that they exhibit 
various kinds of collective behavior.
One possible approach to the mathematical study of such collective behavior 
is to concentrate on certain idealized models 
such as interacting limit cycles and chaotic maps.
An abundance of literature has been devoted to the study of these models
\cite{Winfree,Kuramoto,oscillator1,Kaneko_CML,Kaneko_GCM,Chate_Manneville}.
It is suggested in some foregoing studies 
\cite{Chate_Manneville,Perez_Cerdeira,Pikovsky_Kurths} that
a population of this type as a whole exhibits low dimensional behavior.
This seems to be true even when the individual elements appears to be
mutually uncorrelated,
a situation in which the population could only be 
regarded as a dynamical system of extremely high dimension.
It still remains unclear whether such collective behavior could be understood 
in terms of low-dimensional dynamical systems.

In globally coupled tent maps, collective motion 
has been understood partly from a macroscopic viewpoint 
\cite{Nakagawa_Komatsu,Chawanya_Morita}.
In these studies, the collective motion was discussed only with regard to 
its size, without consideration of its detailed structure.
Some beautiful relations regarding the macroscopic properties 
of the collective motion were explored and a phase diagram was proposed.
These results, however, do not represent progress toward an understanding 
of collective motion in terms of low-dimensional dynamical systems.
In this Paper, 
we first focus on the dynamical properties of collective motion.
We define the Lyapunov exponent characterizing this collective motion
and then a special relation involving this quantity is found.
This relation suggests that the collective motion is always 
chaotic, whenever it appears.
Furthermore, chaotic behavior in the collective motion are found to be 
confined within a small scale whose size is related to the value of 
the Lyapunov exponent.
Finally, on the basis of our numerical results,
we conjecture an answer to the question:
Why does the collective motion appear low dimensional despite the actual 
high dimensionality of the dynamics?

\section{Model : Globally coupled Tent maps}

As model systems, we have chosen globally coupled tent maps, 
because these systems are particularly well suited for detailed numerical 
analysis: We can deal with an ideal limit of infinitely large population.

Globally coupled maps (GCM) are given by an assembly of $N$ elements 
whose behavior is determined by $N$ identical maps with all-to-all coupling. 
The individual elements are then under the influence of a common 
internal field which may be referred to as a mean field \cite{Kaneko_GCM}.
We assume that a single isolated element 
evolves according to $X_{n+1}=f(X_n)$,
where $n$ designates discrete time steps.
Under the interaction through the mean field $h_n$,
the $i$-th element is then assumed to evolve as:
\begin{equation}
X_{n+1}^{(i)}=f(X_n^{(i)})+K h_n, 
\label{eqn:gtent}
\end{equation}
where $K$ is the coupling strength.
In this Paper, we consider the situation in which $f$ is a tent map:
\begin{equation}
f(X)=-a|X|+\displaystyle{{a-1}\over 2},
\label{eqn:tent}
\end{equation}
and the mean field $h_n$ is defined as:
\begin{equation}
h_n \equiv {1\over N}\sum\limits_{j=1}^Nf(X_n^{(j)}).
\end{equation}
Thus, our system is characterized by the two parameters $a$ and $K$
in addition to the total number of elements $N$.
Each tent map has a band splitting point $a=\sqrt{2}$
and we thus stipulate that $a$ satisfies $\sqrt{2}<a<2$.
It is thereby ensured that the population will never split 
into sub-populations.
If the system size $N$ is finite, 
this finiteness becomes the source of fluctuations of the mean field.
Such an effect may obscure pure collective motion.
Thus we work with the limit of large $N$.
In fact, we confirmed that finite size effects can be regarded as noise
acting on the pure collective motion.
For $N\rightarrow \infty$, the population dynamics of GCM 
can be described by the Frobenius-Perron equation \cite{Kaneko3}
for the distribution $\rho (X)$:
\begin{eqnarray}
&&\rho_{n+1}(X)=\int\delta(X-f(X')-Kh_n)\rho_n(X')dX',
\label{eqn:FP}\\
&&h_n=\int f(X')\rho_n(X')dX'.
\label{eqn:op}
\end{eqnarray}
We worked out a numerical scheme for the exact integration of
Eqs.(\ref{eqn:FP}) and (\ref{eqn:op}) whose precision is limited only 
by round-off errors \cite{Morita}. 
Numerical results in this Paper
were obtained using only a single initial distribution, 
uniform over the interval $[f\circ f(0):f(0)]$.

\section{Growth of complexity in collective motion}

Collective motion can be observed by studying the dynamics of the order 
parameter $h_n$ given by Eq.(\ref{eqn:op}).
Figures \ref{fig:snapshot}(a) and \ref{fig:snapshot}(b) are return maps 
for this order parameter.
There we see that
the fluctuations of $h_n$ undergo quasi-periodic motion for small $K$,
but for larger $K$ they display more complicated motion.
Roughly speaking, larger values of $K$ generically result in more complicated 
collective behavior over all values of $a$.

In the case of small $K$, the collective motion appears quasi-periodic.
Despite the familiar appearance of this collective motion,
we should discuss carefully the nature of these dynamics,
because the system has an infinitely high dimension.
The chaotic structure on the torus seems to grow gradually,
without an indication of bifurcation from quasi-periodic to 
the chaotic behavior.
It may be the case that the structure on the torus persists
as long as there is a non-vanishing interaction.
In following sections, 
we focus on the chaotic properties of the collective motion.

\section{Exponential growth of disturbances}

In order to characterize the dynamical properties of the collective motion,
we wish to define a value similar to the Lyapunov exponent to characterize
the collective motion.
We note that it is unfeasible to calculate conventional Lyapunov exponents 
based on the phase space structure of the system,
because our system is of an infinitely high dimension.
However, what we wish to concentrate on is not the phase space structure 
but the dynamical properties of the collective motion.
Hence we attempt to define a quantity similar to the Lyapunov exponent 
to describe the dynamics of the collective motion.

We first add a small disturbance to the distribution $\rho_n(X)$ 
at time step $n$.
The mean field $h_n$ is changed to $h_n^{(d)}$ as a result.
The difference between the disturbed and non-disturbed mean fields,
\begin{equation}
\delta_l \equiv |h_{n+l}^{(d)}-h_{n+l}|,
\label{eqn:delta_l}
\end{equation}
depends on $l$, the time lapsed since the application of the disturbance.
Without an interaction ($K=0$) among the elements in the population,
the distribution $\rho_{n+l}^{(d)}(X)$ gradually approaches
that of the non-disturbed case $\rho_{n+l}(X)$.
This is because the two distributions should realize the same invariant 
measure of a single tent map.
Thus $\delta_l \rightarrow 0$ as $l$ becomes large.
When $K\neq 0$, collective motion generally occurs.
We expect that its dynamical properties can be represented by the behavior of 
$\delta_l$.
If the collective motion is completely quasi-periodic,
the difference $\delta_l$ will remain on the same order as $\delta_0$,
while, if it is chaotic,
$\delta_l$ will grow exponentially with the increase of $l$.

In numerical experiments, $\delta_l$ is estimated for one set of parameters 
$(a,K)$ as follows:
The value $|h_{n+l}^{(d)}-h_{n+l}|$ is calculated
from various initial distributions of $\rho_{n}^{(d)}(X)$
 with a fixed small difference $\delta_0\equiv |h_{n}^{(d)}-h_{n}|$.
The sample average over various $|h_{n+l}^{(d)}-h_{n+l}|$ calculated \
in this way gives $\delta_l$.
In our computations, this average was calculated from either $10$ or $100$ 
initial disturbed distributions.
We point out here that individual $|h_{n+l}^{(d)}-h_{n+l}|$ actually 
depends on $l$ in a manner similar to $\delta_l$.

In Fig.\ref{fig:exp_growth}, we display examples for the exponential growth 
of the difference $\delta_l$ between the mean fields.
The initial difference $\delta_0$ is fixed at as small as $10^{-30}$.
The difference $\delta_l$ exhibits a clear dependence on the time step $l$:
$\delta_l$ grows exponentially and finally saturates near a certain scale.
It is surprising that the exponential growth of $\delta_l$ is observed for a
considerably large interval of $l$.
It is therefore possible to define the exponent $\lambda$ which characterizes 
the exponential growth of $\delta_l$.
The exponent $\lambda$ corresponds to the initial slope 
in Fig.\ref{fig:exp_growth}.
For values of $l$ at which $\delta_l$ displays exponential growth,
we can define $\lambda$ according to
\begin{equation}
\lambda\equiv {1\over l}\log({\delta_l \over {\delta_0}}).
\end{equation}
This exponent was found to be independent of the initial disturbance
of the distribution for sufficiently small values of $\delta_0$ and 
sufficiently large values of $l$.
Hereafter, we call the exponent $\lambda$
the {\it Lyapunov exponent for the collective motion}.

In Fig.\ref{fig:exp_growth} we display three examples of $\delta_l$,
where the values of $K$ are $0.6, 0.8$ and $1.0$ 
while the value of $a$ is fixed according to $-\log_a(2-a)=4$.
Larger values of $K$ results in larger values of $\lambda$.
The collective motion is concluded to be of a chaotic nature
in every case of Fig.\ref{fig:exp_growth}.
If we were to calculate conventional Lyapunov exponents based on the phase 
space structure of this system, a maximal exponent with a value close to 
$\log a$, the Lyapunov exponent for a single tent map, would be obtained.
We note that the value of $\lambda$ is generally smaller than $\log a$.
In the case of Fig.\ref{fig:exp_growth}, $\log a =0.48$, which is larger than
the values found for $\lambda$ (for instance $\lambda=0.17$ when $K=1.0$).
This implies that the dynamics of the collective motion are independent of
the dynamics of the individual elements.

\section{Hilly structure of the Lyapunov exponent}

Figure \ref{fig:a_Ly} shows the Lyapunov exponent $\lambda$
for the collective motion as functions of $a$ ( or, more precisely,
as functions of $m=-\log_a (2-a)$ ) at fixed values of $K$.
Here, hilly structure is displayed by the Lyapunov exponent $\lambda$,
as seen in Fig.\ref{fig:a_Ly}(a).
For comparison, we display the hilly structure corresponding to
{\it macroscopic properties} of the collective motion in Fig.\ref{fig:a_Ly}(b).
These macroscopic properties are represented by the amplitude $F$ of 
the collective motion,
\begin{equation}
F\equiv\sqrt{\langle(h_n-\langle h_n\rangle)^2\rangle}.
\label{eqn:amplitude}
\end{equation}
Here $\langle \rangle$ represents a long time average.
It is worthy to note that the peaks and valleys of these two sets of hilly 
structure seem to be located at the same values of $a$.
For further detail concerning the hilly structure of $F$, 
see \cite{Nakagawa_Komatsu}.
Two special series of parameter values of $a$, the `golden' and `silver'
values, are the key to understanding the hilly structure.
Briefly, the golden values of the parameter $a$ are situated 
in the middle of each hill and the silver values are at the minimum points 
in the valleys.
The same situation is found in the case of the Lyapunov exponent $\lambda$
for the collective motion.

We now concentrate on the differences between these two sets of 
hilly structure.
The hilly structure of the Lyapunov exponent $\lambda$
exhibits gentle ups and downs over the range of values of $a$,
while the amplitude $F$ displays very sharp ups and downs.
The most important difference appears in the valleys of the hilly structure:
The value of $F$ decays linearly to $0$ in each valley, but the value of 
$\lambda$ in each valley seems to remain on the same order as that assumed at
the top of the nearest hill.
However, we will later show that the values of $\lambda$ in fact 
also decay to $0$.
Decaying will be found to occur only in a very narrow region of $a$ 
around the minimum point in each valley.

The regions in which the value of $\lambda$ is close to zero are 
sufficiently narrow that we can ignore them in estimating the order of 
the Lyapunov exponent $\lambda$ as a function of $a$.
For fixed values of $K$ the value of the Lyapunov exponent $\lambda$
decays as $a$ becomes large.
With the definition $m\equiv -\log_a(2-a)$,
this decaying tendency is estimated as
\begin{equation}
\lambda\sim m^{-1},
\label{eqn:Ly_m}
\end{equation}
for positive values of $K$.
In Section VII, we discuss the exceptional regions of $a$ (i.e., those 
in which this relationship does not hold).

\section{Scaling relation of the Lyapunov exponent}

In this section, 
we discuss the relation of the exponent $\lambda$ with the coupling strength 
$K$.
The value of $a$ is confined to be one of the golden values,
while the value of $K$ is changed.
Here a golden value, say $g_p$, is a value of $a$ such that one isolated 
tent maps possesses the property that a trajectory beginning at the peak C 
of the tent map returns to C after $p$ steps.
There exists a sequence of such golden values, parameterized by $p$.
As shown in Fig.\ref{fig:Kvslambda},
it is confirmed that the collective motion becomes increasingly chaotic
as the value of $K$ increases.
For all the golden values we considered, we have found that the Lyapunov 
exponent $\lambda$ grows with the coupling strength $K$ according to 
the relation $\lambda \sim K^2$.
The value of $\lambda$ also obeys the relation Eq.(\ref{eqn:Ly_m}). 
Thus we find that the Lyapunov exponent $\lambda$ for the collective motion 
behaves as
\begin{equation}
\lambda \sim {K^2\over m},
\label{eqn:Ly_K2}
\end{equation}
for the golden values of $a$.
While we have found this relation by confining our consideration to
these golden values, it should be noted that a relation similar to 
Eq.(\ref{eqn:Ly_K2}) is expected to hold for all values of $a$ except in 
very narrow regions,
due to the gentleness of the hilly structure of $\lambda$.
Also note that Eq.(\ref{eqn:Ly_K2}) in some sense describes the envelope
of $\lambda$ in the $(a,K)$ parameter space,
as it does not include the periodic dependence on $m$ or the fine peak
structure of $\lambda$.

Equation (\ref{eqn:Ly_K2}) indicates that when there is an interaction between
elements (i.e. $K\neq 0$) the collective motion is always chaotic.
Thus collective motion displaying quasi-periodicity,
as is seen in Fig.\ref{fig:snapshot}(a), is concluded to actually be chaotic.
As the coupling strength $K$ approaches $0$ or 
the parameter $a$ approaches $2$ (i.e. $m\rightarrow\infty$),
the Lyapunov exponent $\lambda$ for the collective motion vanishes 
asymptotically.

\section{Lyapunov exponents around disappearing points of the collective motion}

The scaling relation for the Lyapunov exponent $\lambda$, Eq.(\ref{eqn:Ly_K2}),
leads us to conclude that the collective motion is always chaotic.
However, there do exist exceptional cases in which the collective motion 
disappears.
In particular, it disappears at the minimum points 
in each valley of the hilly structure of the amplitude $F$.
Around these minimum points,
the Lyapunov exponent for the collective motion is expected to decay 
to vanishing or negative values.
However, it is difficult to detect this behavior in Fig.\ref{fig:a_Ly}.
We now investigate in detail the nature of the Lyapunov exponent 
around the minimum points.

The values of $a$ at the minimum points are called `silver values'.
Here, silver values constitute another particular series of parameter 
values of $a$.
Figure \ref{fig:s_Ly} shows the decaying behavior of the Lyapunov exponent
around a silver value $s_n$ of the parameter $a$.
$|\Delta a|$ represents the distance from $a$ to the silver value $s_n$.
Figure \ref{fig:s_Ly} displays a very narrow neighborhood of $s_n$.
The value of $\lambda$ becomes small as $|\Delta a|$ approaches $0$.
As reflected by this figure, the Lyapunov exponent scales with
the distance $|\Delta a|$ as
\begin{equation}
\lambda\sim (-\log|\Delta a|)^{-{2\over 3}}.
\label{eqn:s_Ly}
\end{equation}
This relation shows that the Lyapunov exponent decays to $0$
as the value of $a$ approach to a silver value.
The Lyapunov exponent is non-positive only at the silver value.
There, the collective motion does not appear, so that chaotic nature of 
this motion is never realized.
What we wish to emphasize is that $\lambda$ falls to zero in only a 
{\it very narrow neighborhood} of the silver value (see Fig.\ref{fig:s_Ly}(b)).
For all other values of $a$, the Lyapunov exponent is approximately
given by Eq.(\ref{eqn:Ly_K2}).
The narrowness of the decaying region around each silver value causes
the apparent smoothness of the Lyapunov exponent as a function of $a$.
The motion of the system is chaotic everywhere, except at the minimum point 
of each valley, where collective motion does not exist.
We now conclude that, whenever collective motion appears (i.e.$F\neq 0$), 
it is chaotic.

\section{minimum structure of the collective motion}

Let us now turn in our investigation of the Lyapunov exponent for the 
collective motion to consideration of the characteristic scale,
that is, the scale $\delta_c$ at which the exponential growth of 
$\delta_l$ saturates (see Fig.\ref{fig:exp_growth}).
The value of $\delta_c$ is determined numerically through the time average 
of $\delta_l$ over an interval of sufficient length just after the saturation.
After the saturation of $\delta_l$, it may continue to increase very slowly,
until it becomes comparable to the size $F$ of the collective motion.
We, however, can choose an appropriate interval for 
obtaining a valid value of $\delta_c$ through the time average 
of $\delta_l$ just after saturation.

Below the scale characterized by $\delta_c$, the collective motion behaves
chaotically corresponding to a linear instability of the collective motion.
The structure characterized by $\delta_c$ is of a scale 
at which the nonlinearity first appears in the collective motion.
The chaotic behavior of the collective motion is confined in this structure.
We hereafter call the structure of the collective motion 
characterized by $\delta_c$ the {\it minimum structure}.
We determined an expression for the characteristic scale $\delta_c$ 
of the minimum structure.
$\delta_c$ was found to depend on the value of $K$ according to
a certain scaling relation.
As is seen in Fig.\ref{fig:2nd_exp}, for sufficiently small values of $K$,
$\delta_c$ behaves as
\begin{equation}
{\delta_c } \sim \exp(-{1\over{K^2}}).
\label{eqn:2nd_exp}
\end{equation}
This figure was obtained at a particular golden value of $a$ 
while the relation (\ref{eqn:2nd_exp}) was confirmed at every golden value 
we considered.
The relation in Eq.(\ref{eqn:2nd_exp}) indicates that 
the minimum structure exists however small the coupling strength $K$ is.

Let us now compare the size $\delta_c$ of the minimum structure with that
of the collective motion.
The amplitude $F$ of the collective motion explored 
in \cite{Nakagawa_Komatsu} obeys a different type of scaling relation,
\begin{equation}
F \sim \exp(-{1\over{K}}).
\label{eqn:1st_exp}
\end{equation}
The scaling relations for both $\delta_c$ and $F$ are given by 
monotonically increasing functions of $K$.
Both sizes decay to $0$ as $K$ decreases.
However, the rate at which this decay occurs is much different for
$\delta_c$ and $F$.
The size $\delta_c$ of the minimum structure 
becomes much smaller than the amplitude $F$ of the collective motion
for sufficiently small values of $K$, 
while they are comparable for $K\simeq 1$.
The minimum structure and the chaotic behavior within it 
belong to a collective set of phenomena in the {\it submacroscopic scale}:
They are properties characterizing the smaller scale of the collective motion,
while the collective motion is a product of the macroscopic nature of 
the distribution $\rho_n(X)$.

\section{Lyapunov exponent and the minimum structure}

For the golden values of the parameter $a$ the size of the minimum structure
is found to be strongly correlated with the value of the Lyapunov 
exponent $\lambda$ for the collective motion:
\begin{equation}
{\delta_c } \sim \exp(-{1\over \lambda}).
\label{eqn:Ly_delta}
\end{equation}
This relation is obtained through the substitution of Eq.(\ref{eqn:Ly_K2}) 
into Eq.(\ref{eqn:2nd_exp}).
We conjecture that this relation is not mere coincidence (i.e., its validity 
is not confined to the golden values) but holds for all sets of $(a,K)$.
The characteristic size $\delta_c$ corresponding to the chaotic behavior
should naturally be related to the Lyapunov exponent.
This conjecture has been confirmed through numerical comparison of
$\lambda$ and $-(\log_a\delta_c)^{-1}$.
In Fig.\ref{fig:Ly_deltac} we show that these two values depend on $a$ 
in a very similar manner, although the values of $-(\log_a \delta_c)^{-1}$ 
are more scattered than those of $\lambda$.
We thus conclude that the relation
\begin{equation}
{\delta_c } \simeq a^{-{1\over {2\lambda}}}
\label{eqn:Ly_delta2}
\end{equation}
is valid for any set of ($a,K$).

The relation in Eq.(\ref{eqn:Ly_delta2}) implies some properties of 
the minimum structure.
As was argued, the Lyapunov exponent $\lambda$ remains 
positive except at the silver values of $a$.
Accordingly, the minimum structure exists for almost all sets of $(a,K)$,
where it becomes small as the chaotic properties become weak.
When the coupling strength $K$ approaches $0$, the values of the 
Lyapunov exponent was found to approach $0$ with the relation in
Eq.(\ref{eqn:Ly_K2}). 
Corresponding to this, the size $\delta_c$ of the minimum structure 
also decreases to $0$ with a much larger decreasing rate with $K$,
as was described in Eq.(\ref{eqn:2nd_exp}).
At the silver values of $a$, the minimum structure does not exist 
($\delta_c=0$), because the collective motion disappears at these points.
The value of $\lambda$ at these points was numerically found to be $0$,
which is consistent with the non-existence of the minimum structure 
($\delta_c=0$).
The relation in Eq.(\ref{eqn:Ly_delta2}) indicates furthermore that
the size $\delta_c$ of the minimum structure also exhibits hilly structure 
as a function of $a$, like $\lambda$ and $F$.
In fact, the hilly structure of $\delta_c$ has been numerically generated.
However, because the values of $\delta_c$ found in this study are widely 
scattered, the functional dependence of $\delta_c$ on $a$ was not made 
completely clear.

It was concluded that the collective motion is almost always chaotic
even though it appears quasi-periodic.
We now obtain a second conclusion from the relation in 
Eq.(\ref{eqn:Ly_delta2}) that the chaotic behavior is always 
confined in a submacroscopic scale given by Eq.(\ref{eqn:Ly_delta2}).
If the coupling strength $K$ is sufficiently small, the chaotic 
behavior is observed only in a much smaller scale than the scale of 
the collective motion.
We argue in a following section that this confinement of the chaotic behavior
leads to the low dimensional appearance of the collective motion.

\section{chaotic fluctuations in the collective motion}

The minimum structure is related to a certain observed structure
in the collective motion.
Consider first the case that the collective motion locks to some periodic 
states, as is seen in Fig.\ref{fig:locking}.
Each periodic point possesses a region which is subject to strong fluctuations
as is shown in the blowups.
The extent of these fluctuations seems to be comparable to the scale $\delta_c$
of the minimum structure.
When, as in Fig.\ref{fig:snapshot}, the collective motion does not lock 
to periodic states, the structure corresponding to $\delta_c$ 
is similar to the case of the locked periodic states.
As is typically observed (see Fig.\ref{fig:snapshot}(b)),
the collective motion fluctuates around some quasi-periodic motion.
The scale of these chaotic fluctuations seems to be on the same order 
as the value of $\delta_c$.

Accepting the correspondence between the extent of the chaotic fluctuations 
and the size of the minimum structure,
we can interpret the observed collective motion 
in terms of the macroscopic properties of the collective motion 
and the minimum structure.
If $K\geq 1$, $\delta_c$ is comparable to the size $F$ of the collective
motion.
In this case the collective motion becomes fully-developed chaotic motion
(see Fig.\ref{fig:snapshot}(b)).
When $K\ll 1$,
the size $\delta_c$ of the minimum structure becomes much smaller than 
the size $F$ of the collective motion.
Here, it is possible for the collective motion to approach a form similar to 
that displayed by a low-dimensional system, as in the case of the 
quasi-periodic motion depicted in Fig.\ref{fig:snapshot}(a).
Our results imply that it is not possible for the system to realize 
quasi-periodic behavior which lacks chaotic fluctuations.
Whenever the collective motion appears, it is always accompanied by 
chaotic fluctuations.
It is important that these chaotic fluctuations are confined within 
the submacroscopic scale whenever they appear.
The collective motion does not realize true quasi-periodic behavior,
however, it reduces to a form which appears to be quasi-periodic 
due to the confinement of the chaotic fluctuations 
within the submacroscopic scale. 
As a result, the quasi-periodic behavior of the collective motion
becomes conspicuous, especially for small values of $K$,
where the characteristic scales of the collective motion and 
the minimum structure are very different.

We now make a conjecture to answer the question of why
the collective motion appears low dimensional despite the actual 
high dimensionality of the dynamics.
Recall now that the collective motion results from dynamics of infinitely high 
dimension.
This is because the present model (Eq.(\ref{eqn:FP}) and 
(\ref{eqn:op})) consists of an infinite number of 
interacting chaotic elements
without any mechanism to reduce the number of the degrees of freedom.
It is natural to imagine that the chaotic fluctuations 
in the submacroscopic scale constitute chaotic behavior of infinitely 
high dimension.
Due to the confinement of the high dimensional chaos within the 
submacroscopic scale, the collective motion reduces to a form 
which appears to be low-dimensional.

\section{Macroscopic behavior vs chaotic fluctuations}

We now use a term {\it macroscopic behavior} definitely for
the coarse-grained behavior of the collective motion 
below the submacroscopic scale,
i.e. below the scale of $\delta_c$ of the minimum structure.
Macroscopic behavior was observed to exhibit low-dimensional motion
such as quasi-periodic and locked periodic motion. 
We now briefly discuss the variety of the macroscopic behavior 
for the collective motion.
The macroscopic behavior of the collective motion changes very frequently
in the parameter space $(a,K)$:
typically the quasi-periodic motion repeatedly locks to and releases from 
some periodic states (see Fig.\ref{fig:locking}).
Examples of the collective motion are displayed in Fig.\ref{fig:nesting}.
The locking phenomena of the collective motion are often followed by
the instability of the locked points.
We find an assembly of tori, as in Fig.\ref{fig:nesting}(a).
These tori can also become locked, where
each torus in the assembly turns into to locked states
so that an assembly of the locked periodic states is formed.
In such a way, a kind of hierarchical structure of locked states 
can be realized when the creation of tori, resulting from the destabilization
of periodic points, is repeated again and again,
leading to the formation of assemblies of tori.
When the collective motion does not lock,
it comes to possess fine structure reflecting a delicate dynamical 
property for the macroscopic behavior (Fig.\ref{fig:nesting}(b)).

As is seen in the above examples, the macroscopic motion can become 
complicated, in some cases possessing fine structure, such as a kind of 
hierarchical structure. 
We here emphasize that, without regard to the dynamical properties of 
the macroscopic scale, the minimum structure inevitably exists
along with the macroscopic behavior of the collective motion.
For all patterns of the macroscopic behavior,
the chaotic behavior only remains below the submacroscopic scale.
In this scale the dynamical properties of the macroscopic behavior is 
destroyed, whereas the chaotic properties in the submacroscopic scale 
do not seem to affect the collective behavior in the macroscopic scale.
The appearance of the minimum structure marks the disappearance of 
the detailed structure of the macroscopic behavior.
The existence of the submacroscopic scale reminds us of the Kolmogorov scale 
in turbulence, where the structure of the energy cascade is destroyed
by the mechanism of thermal dissipation.

\section{Summary and discussion}

In this paper we discussed the collective motion of globally coupled tent maps
from the viewpoint of the dynamical properties of the distribution.
We defined the Lyapunov exponent for the collective motion
from the growth of a disturbance in the distribution $\rho_n(X)$
through projection on the order parameter $h_n$.
The values obtained for the Lyapunov exponent are nontrivial in the sense 
that they cannot be inferred from the dynamical properties of one tent map.
It was found that the collective motion is chaotic whenever it appears,
even though it seems to be periodic or quasi-periodic.
The chaotic behavior of the collective motion seems to exist
inside the minimum structure of the collective motion,
where disturbances grow exponentially.
This minimum structure persists for any collective motion with chaotic 
behavior and its size can be estimated from the value of the Lyapunov exponent.
These results suggest the reason that the collective motion exhibits 
low-dimensional behavior despite the actual high dimensionality.

It is necessary to explain how the minimum structure come into existence.
The scaling relations for the size $\delta_c$ of the minimum structure,
Eq.(\ref{eqn:2nd_exp}), and the Lyapunov exponent $\lambda$ for the 
collective motion, Eq.(\ref{eqn:Ly_K2}), are similar to the forms
obtained analytically by Ershov and Potapov \cite{Ershov_Potapov}.
They considered the size of the fluctuations and Lyapunov exponents
for globally coupled tent maps.
Due to their abstruse analysis, we are not sure if the fluctuations considered
there are related to the minimum structure discussed in this Paper.
We need a more intuitive explanation regarding the occurrence of the chaotic 
fluctuations and their confinement within the minimum structure.


This research is partially supported by JSPS Research Fellowships.

\begin{figure}[h]
\caption{Two types of collective motion at $a=g_5$ 
(defined by $-\log_a(2-a)=4$) for the return map of $h_n$.
(a) For $K=0.1$, we find a torus representing quasi-periodic motion.
(b) For $K=1.0$, the return map displays more complicated torus-like motion,
possibly accompanied by fine structure.
}
\label{fig:snapshot}
\end{figure}

\begin{figure}[h]
\caption{Exponential growth of the difference $\delta_l$ between the 
order parameters $h_{n+l}^{(d)}$ and $h_{n+l}$.
The vertical scale is logarithmic. 
Here $a=g_5$ ($-\log_a(2-a)=4$), while $K=0.6,0.8$ and $1.0$.
All points represent sample averages of $10$ different initial distributions
$\rho_n^{(d)}(X)$ with a fixed difference $\delta_0=10^{-30}$.
}
\label{fig:exp_growth}
\end{figure}

\begin{figure}[h]
\caption{Hilly structure as a function of $m$ for $K=1.0$ ($+$) and 
$0.5$ ($\times$). Here $m=-\log_a(2-a)$. 
(a) $m$ vs Lyapunov exponent $\lambda$. 
In the insert, the same data are displayed on logarithmic scales for both axes.
The dotted line indicates $\lambda=m^{-1}$.
(b) $m$ vs amplitude $F$ of collective motion.
There are several hills and valleys whose locations seem to correspond 
to those in (a).
Some of the golden values of $a$ coincide with integer values of $m$
while some of the silver values coincide with the minimum points of valleys 
of $F$.
}
\label{fig:a_Ly}
\end{figure}

\begin{figure}[h]
\caption{Scaling relation of the Lyapunov exponent $\lambda$ 
for the collective motion with $K$.
Here $\lambda$ vs $K$ is shown for three golden values $g_p$, $p=3$ - $5$.
$a=g_p$ is equal to $-\log_a(2-a)=p-1$.
The dotted line indicates $\lambda=K^{2}$.
}
\label{fig:Kvslambda}
\end{figure}

\begin{figure}[h]
\caption{
Decay of the Lyapunov exponent $\lambda$ for the collective 
motion near a silver value $s_7=1.81327\cdots$. 
Here $\Delta a\equiv a-s_7$.
(a) $-\log|\Delta a|$ vs $\lambda$ with logarithmic scales for both axes.
The dotted line corresponds to $\lambda\propto (-log|\Delta a|)^{-{2\over 3}}$.
(b) $|\Delta a|$ vs $\lambda$ with normal scales for both axes. 
The region where $\lambda$ decays to $0$ is found to be very narrow 
when $\lambda$ is plotted in this way.
}
\label{fig:s_Ly}
\end{figure}

\begin{figure}[h]
\caption{Scaling relation of the size $\delta_c$ of the minimum structure 
with $K$.
Here $\delta_c$ vs $1/K^2$ is shown for the golden value 
$a=g_4$ ($-\log_a(2-a)=3$).
The vertical scale is logarithmic.
The dotted line indicates $\delta_c \propto \exp(-K^{-2})$.
}
\label{fig:2nd_exp}
\end{figure}

\begin{figure}[h]
\caption{Relation between the Lyapunov exponent $\lambda$ and 
the size $\delta_c$ of the minimum structure.
Here $m=-\log_a(2-a)$. This figure is for the case $K=0.5$.
(a) The Lyapunov exponent $\lambda$ for the collective motion.
(b) $\lambda^{\prime}\equiv (-2\log_a\delta_c)^{-1}$, where $\delta_c$ is the 
size of the minimum structure (also see Fig.\protect{\ref{fig:exp_growth}}).
We find good correspondence between (a) and (b).
}
\label{fig:Ly_deltac}
\end{figure}

\begin{figure}[h]
\caption{Locking phenomenon of the collective motion at $a=g_4$
($\log_a(2-a)=3$) and $K=0.5$, for the return map of $h_n$.
In the blowups, fluctuations are found around each locked point.
The extent of the fluctuations is comparable to the size $\delta_c$ of 
the minimum structure at a corresponding set of parameters $(a,K)$.
}
\label{fig:locking}
\end{figure}

\begin{figure}[h]
\caption{Examples of collective motion. Here $a=g_4$ ($-\log_a(2-a)=3$).
Each figure is for the return map of $h_n$.
(a) $K=0.501$. Each periodic point in Fig.\protect{\ref{fig:locking}}
becomes a torus due to the instability of the locked point.
For a slightly larger value of $K$, these small tori exhibit period locking 
again.
(b) $K=0.507$. The locked state is not stable here. 
A kind of delicate structure is observed, reflecting the stability of the
macroscopic dynamics.
Points are seen to be scattered when a pattern is observed at a sufficiently 
small scale.
These fluctuations appear even after a transient behavior has disappeared.
}
\label{fig:nesting}
\end{figure}

\end{document}